\documentclass[aps,floatfix,showpacs,twocolumn]{revtex4}
\usepackage{graphicx}
\usepackage{epsfig}

\begin{document}

\title{On perfect fluid models in non-comoving observational spherical
  coordinates} 
\author{Mustapha Ishak}
\email{mishak@princeton.edu}

\affiliation{
Department of Astrophysical Sciences and Department of Physics, Princeton University, Princeton, NJ 08544, USA}
\date{\today}

\begin{abstract}
We use null spherical (observational) coordinates to describe a class
of inhomogeneous cosmological models. The proposed cosmological
construction is based on the observer past null cone. A known
difficulty in using inhomogeneous models is that the null geodesic
equation is not integrable in general. Our choice of null coordinates
solves the radial ingoing null geodesic by construction. Furthermore,
we use an approach where the velocity field is uniquely calculated
from the metric rather than put in by hand. Conveniently, this allows
us to explore models in a non-comoving frame of reference. In this
frame, we find that the velocity field has shear, acceleration and
expansion rate in general. We show that a comoving frame is not
compatible with expanding perfect fluid models in the coordinates
proposed and dust models are simply not possible. We describe 
the models in a non-comoving frame. We use the dust models in a 
non-comoving frame to outline a fitting procedure.
\end{abstract}


\maketitle
\section{Introduction}
The study of inhomogeneous cosmological models is a well motivated and
justified endeavor (see for reviews \cite{krasinski,Molina}). These
models provide more freedom 
in discussing very early or very late evolution of the irregularities
in the Universe. Their study also complements perturbation approaches. 
It is worth mentioning that there are a few hundreds of inhomogeneous
cosmological models that reproduce a metric of the
Friedmann-Lemaitre-Robertson-Walker (FLRW) class of solutions when
their arbitrary constants or functions take certain limiting
values \cite{krasinski}. They become then, in that limit, 
compatible with the almost homogeneous and almost isotropic 
observed Universe. This shows the richness of these studies. 

A difficulty that is encountered in this models is that the null
geodesic equation is not integrable in general. In this paper, we
explore the alternative of using null (observational) spherical
coordinates in which the radial null geodesic equation of interest is
solved by construction. 
However, when considering null coordinates and a given metric for the 
spacetime some subtleties arise regarding the frame of reference used. 
In order to explore this point we will use in this paper the approach 
described in Ishak and Lake\cite{IshakLake2003} where the velocity 
field is calculated from the metric and not put in by hand. 
Conveniently, this approach allows one to explore non-comoving frames 
of reference, an important point for this paper.  

Surprisingly, little work has been done in non-comoving coordinates
\cite{narlikar1,mogue,vaidya1968,mcvittie,Senovilla,Davidson,kshm}
despite some interesting features particular to them. Notably, there
are models which are separable only in a non-comoving coordinate system
\cite{Senovilla}. Moreover, exact solutions to Einstein's equations in a
non-comoving frame usually have a rich kinematics with shear,
acceleration and expansion. Such solutions are relatively rare in the
comoving frame \cite{kshm}, see also a recent discussion in \cite{Davidson}. 
Another point discussed in \cite{Senovilla} is that comoving
coordinates do not cover all the spacetime manifold for a specified
energy momentum tensor. Finally, it is worth mentioning that it
is often difficult to do the mathematical transformation of a given
solution from non-comoving coordinates to comoving ones, and even when
the passage is made, there is no guarantee that the solution will
continue to have a simple or explicit form.  

In the present paper, we use the spherical null Bondi metric \cite{bondip}
to present models where a non-comoving frame is proven necessary.
We explicitly demonstrate how a comoving frame leads to severe limitations. 

Furthermore, we use the dust models in the non-comoving frame to outline 
a fitting procedure where observational data can be used to integrate 
explicitly for the metric functions.
Using observational coordinates is particularly useful 
when one wants to compare directly an inhomogeneous model to observational
data. Such an interesting program had been nicely developed in Refs.  
\cite{ellisdel,stoegerellis2,stoegerellis,stoegerellis3,stoegerellis4,stoegerellis5} 
where the authors used a general metric that can be written as 
a FLRW metric plus exact perturbations. 
The spherically symmetric dust solutions were considered in 
Ref. \cite{stoegerellis}. The authors also developed 
and used a fluid-ray tetrad formalism \cite{stoegerellis2} in order 
to derive a fitting procedure where observations can be used to 
solve the Einstein's field equations. 
After some necessary revisions \cite{maartens,araujo1},
this program has been relaunched recently \cite{araujo1,araujo2}.

We consider here in our work the spherically symmetric case but using
the Bondi metric \cite{bondip} in a non-comoving frame.
Also, we don't use the fluid-ray tetrad formalism  \cite{stoegerellis2} but the 
inverse approach to Einstein's equations developed in \cite{IshakLake2003}.

In the following section, we set the notation and
recall some useful results. In section III, we discuss observational
coordinates and explain the cosmological construction around our
world-line. We also discuss the physical meaning of the functions that
appear in the metric used here. We provide in section IV perfect fluid
models in a non-comoving frame. In section V, we show how dust models
are not possible in a comoving frame. We describe dust models in a 
non-comoving frame  and outline a fitting procedure in section VI 
and summarize in Section VII.
\section{Notation and preliminaries}
We set here the notation and summarize
results to be used in this paper. In Ref. \cite{IshakLake2003}, warped product
spacetimes of class $B_{1}$   
\cite{santo,carot} were considered. These can be written in the form
\begin{equation}
ds^2_{\mathcal{M}}=ds^2_{\Sigma_{1}}(x^1,x^2)+C(x^{\alpha})ds_{\Sigma_{2}}^{2}(x^3,x^4)\label{gmetric}
\end{equation}
where $C(x^{\alpha})=r(x^1,x^2)^2 w(x^3,x^4)^2$,
$sig(\Sigma_{1})=0$ and
 $sig(\Sigma_{2})=2 \epsilon$ $(\epsilon=\pm1)$.
Although very special, these spaces include many of interest, for
example, \emph{all} spherical, plane, and hyperbolic spacetimes.
For $\Sigma_{1}$, we write
\begin{equation}
ds^{2}_{\Sigma_{1}}=a(dx^1)^{2}+2bdx^1dx^2+c(dx^2)^{2},
\label{metric}
\end{equation}
with $a,b$ and $c$ functions of $(x^1,x^2)$ only. 
Consider a congruence of unit timelike vectors (velocity field)
$u^{\alpha}=(u^1,u^2,0,0)$ with an associated unit normal field
$n^{\alpha}$ (in the tangent space of $\Sigma_{1}$) satisfying
$n_{\alpha}u^{\alpha}=0, n_{\alpha}n^{\alpha}=1$ \cite{notation}. It
was shown in \cite{IshakLake2003} that $u^{\alpha}$ is uniquely
determined from the zero flux condition 
\begin{equation}
G_{\alpha}^{\beta}u^{\alpha}n_{\beta}=0, \label{zeroflux}
\end{equation}
where $G_{\alpha}^{\beta}$ is the Einstein tensor of
the spacetime.
The explicit forms for $u^1$ and $u^2$ were written out for canonical
representations of $\Sigma_{1}$, including the
null (Bondi) type of coordinates that we use in the present
paper. With $G\equiv G^{\alpha}_{\alpha}$, $G1 \equiv
G_{\alpha}^{\beta}u^{\alpha}u_{\beta}$ and $G2 \equiv
G_{\alpha}^{\beta}n^{\alpha}n_{\beta}$, it was shown in
\cite{IshakLake2003} that the condition  
\begin{equation}
G+G1=3G2\label{giso}
\end{equation}
is a necessary condition for a perfect fluid source,
and that in some cases, this condition is also 
sufficient. For example, in \cite{Lake2003}, equation (\ref{giso}) was
used to derive an algorithm which generates all regular static
spherically symmetric perfect fluid solutions of Einstein's equations.
In this paper, we are interested in perfect fluid sources so it is
important to recapitulate the following results from
\cite{IshakLake2003}. Consider a fluid with anisotropic pressure and
shear viscosity but zero energy flux (non-conducting). The
energy-momentum reads   
\begin{equation}
T^{\alpha}_{\beta}=\rho
u^{\alpha}u_{\beta}+p_{1}n^{\alpha}n_{\beta}
+p_{2}\delta^{\alpha}_{\beta} +
p_{2}(u^{\alpha}u_{\beta}-n^{\alpha}n_{\beta})-2\eta
\sigma^{\alpha}_{\beta}, \label{imperfect}
\end{equation}
where $\rho$ is the energy density and $\sigma^{\alpha}_{\beta}$ is
the shear associated with $u^{\alpha}$; $\eta$ is the phenomenological
shear viscosity; $p_1$ and $p_2$ are the pressures respectively
parallel and perpendicular to $n^{\alpha}$. When $p_1=p_2$
and the shear term vanishes the fluid is called perfect. It was
shown in \cite{IshakLake2003} that in the case where 
\begin{equation}
\Delta\equiv \sigma_{\alpha}^{\beta}n^{\alpha}n^{\beta}\ne 0,\label{Delta}
\end{equation}
we have 
\begin{equation}
\rho=\frac{G1}{2\pi},\label{energydensity}
\end{equation}
\begin{equation}
p1=\frac{G2}{8 \pi}+2 \eta \Delta, \label{p1s}
\end{equation}
\begin{equation}
p2=\frac{G+G1-G2}{16 \pi}-\eta \Delta \label{p2s}
\end{equation}
and $\eta$ is a freely specified function. The procedure to impose a
perfect fluid source in this degenerate case is to impose the condition
(\ref{giso}) and also necessarily set $\eta\equiv 0$. For other
 choices of $\eta$, the fluid is viscous.
\section{The metric and observational coordinates}\label{metricsection}
\begin{figure}
\centerline{\psfig{file=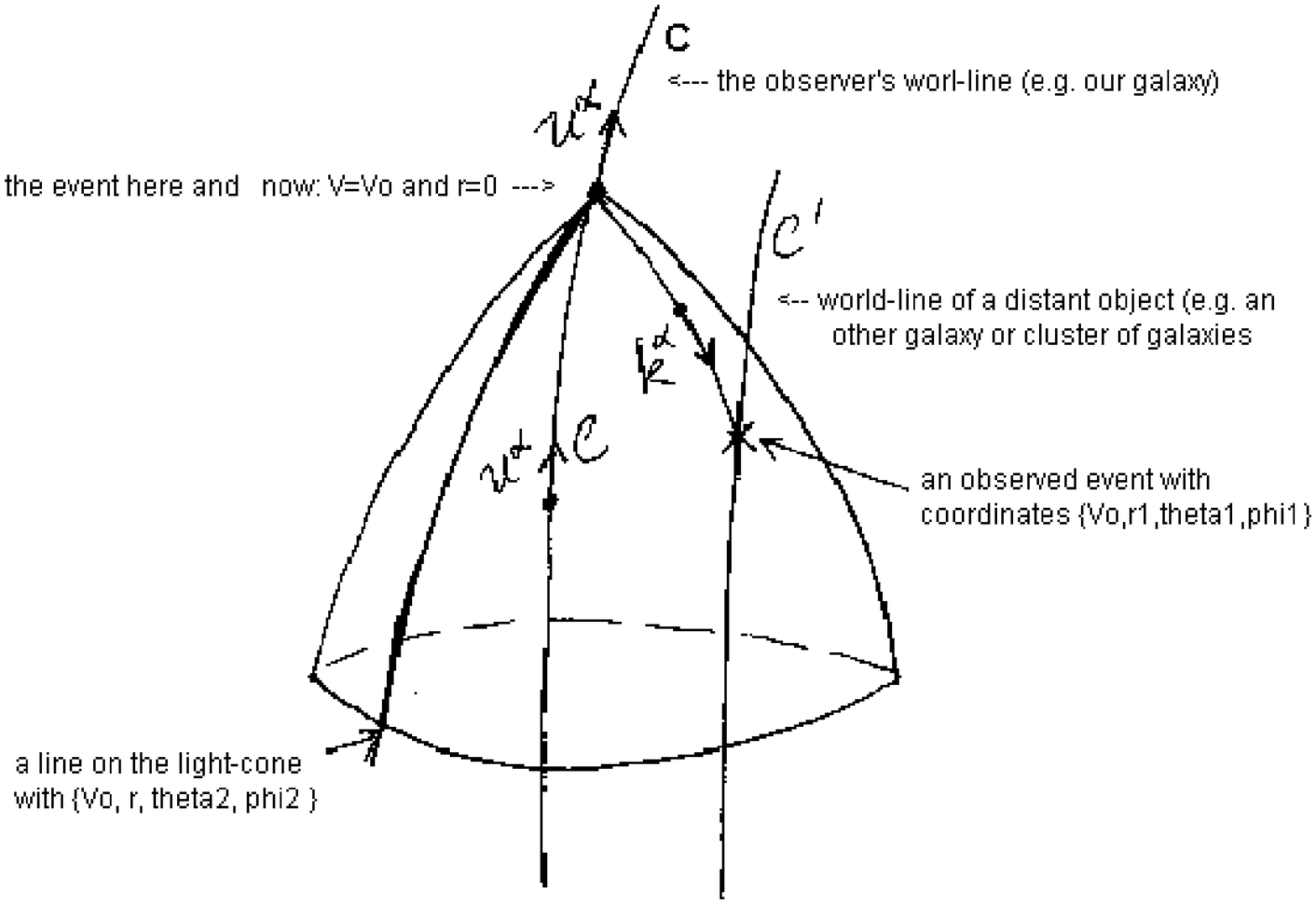,width=10 cm,height=8 cm}} 
\caption
{Observational coordinates $\{v,r,\theta,\phi\}$. 
Our past null cone is defined by $v=v_{o}$. C is the  observer
  world-line, C' is the world-line of another celestial object. The
  trajectory defined by $v$, $\theta$ and $\phi$ constant is a radial
  null geodesic, $u^{\alpha}$ is the fluid velocity field
  vector and $k^{\alpha}$ is the null tangent vector. The null
  rays are traveling in the opposite direction to $k^{\alpha}$ as
  drawn on the figure. The coordinate $r$ increases along the
  trajectory CC' down the light-cone. 
  }
\label{coordinates}
\end{figure}

We consider in the present paper the null coordinate system
$\{x^a\}=\{v,r,\theta,\phi\}$. These are called observational 
(or cosmological) coordinates as we can construct them around our
galaxy world line C as indicated in  Figure \ref{coordinates}. 
The trajectory defined by $v$, $\theta$ and $\phi$ constant is a radial
null geodesic and each hyper-surface of constant $v$ is a past light
cone of events on C. We choose $v=v_o$ and $r=0$ to represent the
vertex ``here and now''. The coordinate $r$ is then set by
construction to be the area distance as explained further and is
related to the luminosity distance $d_{L}$ by $r=d_{_{L}}/(1+z)^2$ (see
e.g.\cite{gfre}). Finally, $\theta$ and $\phi$ are the spherical
coordinates on the celestial sphere. The geometry of the models is
represented by the general spherical Bondi metric in advanced
coordinates \cite{bondip}:  
\begin{equation} 
{ds^2 = 2\, c\, dr\, dv - c^2 (1 - \frac{2 m}{r}) dv^2 + r^2
(d\theta^2 + sin(\theta)^2 d\phi^2)} \label{metric} 
\end{equation}    
where $c\equiv c(r,v)>0$, $m\equiv m(r,v)$ and $r>2m$.  
The radial ($\theta$ and $\phi$ constant) ingoing null geodesic
equation $v$=constant is solved by construction (see Appendix
\ref{appNull}). The components of the mixed Einstein tensor
$G^{\alpha}_{\beta}$ for (\ref{metric}) are 
given in appendix \ref{appEinstein} and the structure of the Weyl
tensor is discussed in Appendix \ref{appWeyl}.
Regularity of the metric and the Weyl invariants 
requires that $m(r,v)$ and $c(r,v)$ are $C^3$ at $r=0$ (e.g. see
equation (\ref{W})). It follows that 
\begin{equation}
(1-\frac{2m(r,v)}{r})\Big{|}_{r=0}=1.\label{regul}
\end{equation}  
Also, we can use the freedom in the null coordinate $v$ to  normalize it by
setting $c(0,v)=1$. As we will write further in this paper (see equation
(\ref{normalization})) this means that we require that $v$ measures the
proper time $\tau$ along our galaxy world line C. 

Whereas the meaning of the metric function $m(r,v)$ is very well known, 
we are not aware of any previous literature where an interpretation
for $c(r,v)$ was given. The function $m(r,v)$ represents the 
effective gravitational (geometrical) mass (e.g.
\cite{mass1}, \cite{mass2}, \cite{mass3}, \cite{mass4} and
\cite{mass5}) and is given by 
\begin{equation}
{m \equiv \frac{g_{\theta \theta}^{3/2}}{2} R_{\theta
\phi}{}^{\theta \phi}}\label{effectivemass} 
\end{equation}
where $R_{\theta \phi}{}^{\theta \phi}$ is the mixed angular component
of the Riemann curvature tensor.
For the physical meaning of the function $c(r,v)$, it turns out
to be useful to study the kinematics of null rays. These usually 
include the optical shear, vorticity and rate of expansion,
respectively defined by \cite{sachs}  
\begin{eqnarray}
\sigma_{optical}^{2} & \equiv & \frac{1}{2}k_{(\alpha;\beta)}k^{\alpha;\beta}-\frac{1}{4}
(k^{\alpha}_{;\alpha})^2, \\
\omega_{optical}^{2} & \equiv&  
\frac{1}{2}k_{[\alpha;\beta]}k^{\alpha;\beta},  \\
\theta_{optical} &   \equiv &  \frac{1}{2}k^{\alpha}_{;\alpha} \label{opticalexpansion}
\end{eqnarray}
where 
\begin{equation}
k_{\alpha}=[1,0,0,0]
\label{nullvector} 
\end{equation}
is the null 4-vector tangent to the
congruence of null geodesics. The physical meaning of the
optical scalars can be understood in the following way
\cite{sachs,DInverno}:  
If an opaque object is displaced an infinitesimal distance $dr$ from a 
screen (perpendicularly to the beam of light), it will cast on the
screen a shadow that is expanded by $\theta_{optical}dr$, rotated by
$\omega_{optical}dr$ and sheared by $|\sigma_{optical}|dr$.
As expected from the spherical symmetry of the geometry,
the non-vanishing optical scalar for the null congruence $k_{\alpha}$
is the optical rate of expansion, from which we find
\begin{equation}
rc(r,v) = \frac {1}{\theta_{optical}}\label{cvalue}
\end{equation}
We identify from equation (\ref{cvalue}) that $rc(r,v)$ is a measure
of the reciprocal of the expansion of null rays.
\section{Models in a non-comoving frame}
\subsection{The velocity field}
We consider an observer moving with a fluid for which the streamlines
are given by the general radial timelike vector
$u^{\alpha}=[u^{v}(r,v),u^{r}(r,v),0,0]$. We assume that such a
velocity field exists for which 
the energy-momentum tensor takes the perfect fluid form   
\begin{equation}
T^{\alpha}_{\beta}=(\rho+p)
u^{\alpha}u_{\beta}+p\delta^{\alpha}_{\beta}
\end{equation}
where $\rho$ and $p$ are respectively the energy-density and isotropic
pressure associated with $u^{\alpha}$. The velocity field is simply
determined from the zero flux condition (\ref{zeroflux})
and is given by 
\begin{equation}
u^{1} \equiv
u^{v}=\frac{1}{c(r,v)}\sqrt[4]{\frac{1}{\big{(}1-\frac{2m(r,v)}{r}\big{)}^2
    +\frac{4m^{\bullet}(r,v)}{rc'(r,v)}}} 
\label{u1b}
\end{equation}
\begin{equation}
u^{2} \equiv
u^{r}=\frac{1}{2}
\frac{\big{(}1-\frac{2m(r,v)}{r}\big{)}
  -\sqrt(\big{(}1-\frac{2m(r,v)}{r}\big{)}^2+\frac{4m^{\bullet}(r,v)}{rc'(r,v)})}
{\sqrt[4]{\big{(}1-\frac{2m(r,v)}{r}\big{)}^2+\frac{4m^{\bullet}(r,v)}{rc'(r,v)}}}  
\label{u2b}
\end{equation}
where $'\equiv \frac{\partial}{\partial r}$ and ${}^{\bullet}\equiv
\frac{\partial}{\partial v}$. The associated unit normal vector field 
$n^{\alpha}$ ($n_{\alpha}u^{\alpha}=0$ and 
$n_{\alpha}n^{\alpha}=1$) is given by 
\begin{equation}
n_{\alpha}=c(r,v)[u^{r},-u^{v},0,0].
\end{equation}

Interestingly, the velocity field has shear $\sigma^{\alpha}_{\beta}
\not = 0$, acceleration $\dot{u}^{\alpha} \not =0$ and
expansion rate scalar $\theta \not = 0$ in general.
\subsection{The perfect fluid condition}
It follows from the metric (\ref{metric}) that $\Delta$, as defined in
  (\ref{Delta}), is not zero and that for a perfect fluid source we
  must impose the condition (\ref{giso}) and set $\eta\equiv
  0$. With $m\equiv m(r,v)$ and $c\equiv c(r,v)$, condition
  (\ref{giso}) reads 
\begin{equation}
\mathcal{L}-c^2\sqrt{c'\mathcal{N}}=0\label{bondipfcondition}
\end{equation}
where 
\begin{eqnarray}
\mathcal{L}&\equiv& c^3(2m'-m''r)+c^2(3c'(m-rm')\\ \nonumber
 & &+c''r(r-2m))+cr^2c'{}^{\bullet}-c'r^2c^{\bullet}
\end{eqnarray}
and
\begin{equation}
\mathcal{N}\equiv c'(r-2m)^2+4rm^{\bullet}.
\end{equation}
The metric (\ref{metric}) along with the metric constraint
(\ref{bondipfcondition}) represent a perfect fluid model with
\begin{equation}
\rho  = \frac{G1}{8 \pi}={\frac{2(cm)'-c'r+\sqrt{c'\mathcal{N}}}{8
    \pi cr^2}}\label{rhob}
\end{equation}
and 
\begin{equation} 
p \; (=p_1=p_2) = {\frac{-2(cm)'+c'r+\sqrt{c'\mathcal{N}}}{8 \pi cr^2} }\label{p1b}.
\end{equation}
\section{Models in a comoving frame}
In this section, we specialize to models in a comoving
frame of reference. We show how this frame fails in the
realization of the cosmological construction proposed.
\subsection{Perfect fluid models} 
With the metric function 
\begin{equation}
g_{vv}= -c^2(r,v)\Big{(}1-\frac{2m(r,v)}{r} \Big{)} \; < \; 0,
\end{equation}
the requirement of comoving coordinates reads
\begin{equation}
u^{r}=0 \; \; \Leftrightarrow \; \; G^{r}_{v}=2 \frac{\frac{\partial
    m(r,v)}{\partial v}}{r^2}=0.
\end{equation}
Hence, the necessary and sufficient condition for a comoving frame
is
\begin{equation}
  m(r,v)=m(r).\label{comovdustconst1}
\end{equation}
It follows from (\ref{u1b}) and (\ref{comovdustconst1}) that
\begin{equation}
u^{v}=\frac{1}{c(r,v)\sqrt{\big{(}1-\frac{2m(r)}{r} \big{)} }},
\end{equation}
\begin{equation}
n_{r}=\frac{-1}{\sqrt{\big{(}1-\frac{2m(r)}{r} \big{)} }} 
\end{equation}
and $n_{v}=0$.
With this velocity field the shear tensor vanishes, therefore, the
necessary and sufficient condition for a perfect fluid model is
equation (\ref{giso}), which can be written as   
\begin{eqnarray}
-c^2c'r+5c^2c'm+2c^3m'+c^2c''r^2-2c^2c''mr-& \nonumber\\
3c^2c'm'r-c^3m''r-c'c^{\bullet}r^2+c'{}^{\bullet}cr^2=0.&\label{dustpfc} 
\end{eqnarray} 
For a perfect fluid source in this frame the pressure $p$ is a
function of both $r$ and $v$ while the energy density is only a
function of $r$,  
\begin{equation}
\rho=\frac{G1}{8\pi}=\frac{m'(r)}{4 \pi r^2}
\end{equation}
and
\begin{equation}
p=\frac{c'(r,v)(r+2m(r))+c(r,v)m'(r)}{4 \pi r^2 c(r,v)}.
\end{equation}
The 4-acceleration $\dot{u}^{\alpha}$ has the non-vanishing components
\begin{equation}
\dot{u}^{v}=\frac{rc'(r,v)(r-2m(r))+c(m(r)-m'(r)r)}{rc^2(r,v)(r-2m(r))}  
\end{equation}
and
\begin{equation}
\dot{u}^{r}=\frac{rc{'}(r,v)(r-2m(r))+c(m(r)-m'(r)r)}{r^2c(r,v)}.
\end{equation}
A caveat in this frame (comoving) is that the expansion scalar vanishes and the
model is not suitable for describing an expanding Universe.
\subsection{The zero-pressure case}
The present matter dominated (as opposed to radiation dominated) Universe
 is well approximated by a zero-pressure model, commonly referred
to as ``dust''. In this case (comoving) $\Delta=0$ and the zero
 pressure conditions follow from equations (\ref{p1s}) and
 (\ref{p2s}) as
\begin{equation}
G2= 0\label{G2bondicomovdust}
\end{equation}
and
\begin{equation}
G+G1=0. \label{GG1bondicomovdust}
\end{equation}
With (\ref{comovdustconst1}), equations (\ref{G2bondicomovdust}) and
(\ref{GG1bondicomovdust})  read
\begin{equation}
\frac{c'}{c}-\frac{m'}{r-2m}=0,\label{G2bondicomovdust2}
\end{equation}
and
\begin{eqnarray}
c^{2}c'r+c^2 c'm+c^{2}
c''r^2-2c^{2}c''rm-3c^{2}c'rm'- & \nonumber \\
c^{3}m''r-c'c^{\bullet}r^2+c'^{\bullet}cr^2=0.&
\label{GG1bondicomovdust2} 
\end{eqnarray}
Integrating (\ref{G2bondicomovdust2}) gives
\begin{equation}
c(r,v)=f(v)\exp\Big{(}\int{\frac{m'(r)}{r-2m(r)}dr}\Big{)}\label{sol1}
\end{equation}
which when put into (\ref{GG1bondicomovdust2}) gives 
\begin{equation}
\frac{f(v)^3\exp\Big{(}3\int{\frac{m'(r)}{r-2m(r)}dr}\Big{)} m'(r)m(r)}{r^2(r-2m(r))}=0.
\end{equation}
With  $f(v)>0$ (from $c(r,v)>0$) the zero pressure model reduces  
to the following two cases:
\\ 
i) $m(r)=0$ and the spacetime reduces to the
Minkowski flat  spacetime ($R_{\alpha\beta\gamma\delta}=0$), or,
\\ 
ii)$m'(r)=0$  ($m$ is constant) and the spacetime reduces to
Schwarzschild vacuum in Eddington-Finkelstein coordinates
($R_{\alpha\beta}=0$, $R_{\alpha\beta\gamma\delta}\not=0$.)
\\
Therefore, a dust model is not possible in a comoving frame using the
observational coordinates and the Bondi metric (\ref{metric}). We turn in the
following section to a non-comoving frame for dust models.
\section{Dust models in a non-comoving frame}
\subsection{The velocity field}
We are interested in building dust models using spherical observational
coordinates and a non-comoving frame. In a 1+3 decomposition of the
spacetime, these models are given by the general Lema\^{\i}tre-Tolman-Bondi
solution \cite{lemaitre}\cite{krasinski}. In this non-comoving case
$\Delta\ne 0$ in general, so we must set $\eta\equiv 0$ and impose the
zero pressure conditions (\ref{G2bondicomovdust}) and
(\ref{GG1bondicomovdust}) which can be written as 
\begin{equation}
{m^{\bullet}=cm'\Big{(}\frac{c
m'}{c' r} - \Big{(}1 - \frac{2m}{r}\Big{)}\Big{)}}.\label{dustconst1}
\end{equation}
and
(\ref{GG1bondicomovdust}) can be written as 
\begin{equation}
c'{}^{\bullet}=c\Big{(}\frac{1}{r}(3c'm'+cm'')-\frac{c'}{r}(1+\frac{m}{r})-c''(1-\frac{2m}{r})\Big{)}+\frac{c'c^{\bullet}}{c}\label{dustconst2}   
\end{equation}
The metric (\ref{metric}) with constraints (\ref{dustconst1}) and
(\ref{dustconst2}) represents a class of inhomogeneous dust
models in spherical observational coordinates. Using equation
(\ref{energydensity}), the energy density is given by
\begin{equation}
4 \pi
\rho(r,v)= \frac{2m'(r,v)}{r^2}-\frac{c'(r,v)}{rc(r,v)}\Big{(}1-\frac{2m(r,v)}{r}\Big{)}.
\label{dustrhomass}
\end{equation} 
This result can also be obtained from the effective
gravitational mass equation (\ref{effectivemass}). 
The velocity field follows from equations (\ref{u1b}) and (\ref{u2b}):
\begin{equation}
{u^{v}=\frac{1}{c(r,v)}\frac{1}{\sqrt{(1-\frac{2m(r,v)}{r})+\frac{2m^{\bullet}(r,v)}{m'c}}}}
\label{u1bd1}  
\end{equation}  
or equivalently by using (\ref{dustconst1})
\begin{equation}
{u^{v}=\frac{1}{c(r,v)}\frac{1}{\sqrt{\frac{2m'(r,v)c(r,v)}{rc'(r,v)}
-(1-\frac{2m(r,v)}{r})}}} \label{u1bd2} 
\end{equation} 
and 
\begin{equation}
u^{r}=\frac{m^{\bullet}(r,v)}{m'(r,v)}u^{v}.
\end{equation}  
We verified that the acceleration 4-vector field $\dot{u}^{\alpha}$ vanishes 
as the dust fluid is moving geodesically. Interestingly, the velocity field
remains with non-vanishing shear and expansion rate.
\subsection{The conformally flat case}
It is a well known result that a cosmological model which satisfies the
Einstein equations with a perfect fluid source, a barotropic
equation of state i.e. $p=p(\rho)$ (including $p=0$), which is
conformally flat ($C_{\alpha\beta\gamma\delta}=0$) and has non-zero
expansion is a Lemaitre-Friedmann-Robertson-Walker model (LFRW)
\cite{krasinski}, \cite{kshm} and \cite{stephani}.  
For dust models in the non-comoving frame, the condition
$C_{\alpha\beta\gamma\delta}=0$ (see appendix \ref{appWeyl}) reduces to  
\begin{equation}
\frac{c'}{c}\Big{(}1-\frac{2m}{r}\Big{)}-\frac{2m'}{r}+\frac{3m}{r^2}=0\label{cflatdust}
\end{equation}
Therefore, the metric (\ref{metric}) along with the constraints
(\ref{dustconst1}), (\ref{dustconst2}) and (\ref{cflatdust})
represents the homogeneous and isotropic (LFRW) limit of the
models. We are interested here in more general inhomogeneous
models.
\subsection{Basic observable quantities}
\subsubsection{The Redshift}
The light emitted with a wavelength $\lambda_{e}$ from a point
on the light cone is observed at the vertex ``here and now'' (see
Figure \ref{coordinates}) with a wavelength $\lambda_{o}$. The
redshift is  then given by (see e.g. \cite{krisac}, \cite{gfre}) 
\begin{equation}
{1+z= \frac{(k_{\alpha} u^{\alpha})_{emitter}}{(k_{\beta}
    u^{\beta})_{observer}}=\frac{d\tau_{observer}} 
{d\tau_{emitter}}=\frac{\lambda_o}{\lambda_e}}
\end{equation} 
where $u^{\alpha}$ is the normalized timelike velocity vector field and
$k_{\alpha}$ is the null vector as given previously by
(\ref{nullvector}). It follows that  
\begin{equation}
k_{\alpha} u^{\alpha}= u^{v}(r,v_o)
\end{equation}
where $u^{v}(r,v_o)$ is given by (\ref{u1b}).   
It follows from the regularity
condition (\ref{regul}) and the timelike normalization condition
$u^{\alpha}u_{\alpha}=-1$ evaluated at $r=0$ (observer) that  
\begin{equation}
(k_{\alpha} u^{\alpha})\Big{|}_{observer}=
  u^{v}(0,v_o)=\frac{1}{c(0,v_o)}=1 \label{normalization}
\end{equation}
where in the last step we used the freedom in the null coordinate $v$ to
set $c(0,v)=0$. Finally,
\begin{equation}
{1+z=\frac {u^{v}(r,v_o)_{emitter}}{u^{v}(r,v_o)_{observer}}=u^{v}
(r,v_o)_{emitter}}.\label{redshift1}
\end{equation}   
For the dust case, equation (\ref{redshift1}) gives
\begin{equation}
{1+z=\frac{1}{c(r,v_o)}\frac{1}{\sqrt{\frac{2m'(r,v_o)c(r,v_o)}{rc'(r,v_o)}
-(1-\frac{2m(r,v_o)}{r})}}} \label{dustredshift1} 
\end{equation}  
or equivalently by using the constraint (\ref{dustconst1})
\begin{equation}
{1+z=\frac{1}{c(r,v_o)}\frac{1}{\sqrt{\frac{m'(r,v_o)c(r,v_o)}{rc'(r,v_o)}+
\frac{{m}^{\bullet}(r,v_o)}{m'(r,v_o)c(r,v_o)}}}}
\label{dustredshift2}   
\end{equation}
where we have set $r \equiv r_{emitter}$ in equations
(\ref{dustredshift1}) and (\ref{dustredshift2}).
\subsubsection{The observer area distance}
The coordinate $r$ in the model is set by construction to be the
observer area distance  \cite{gfre} \cite{krisac} which is defined by 
$dA=r^{2}d\Omega$ where $dA$ is  
the cross-sectional area of an emitting object, and $d\Omega$ is the
solid angle subtended by that object at the 
observer. The area distance $r$ is related to the
luminosity distance $d_{L}$ by $r=d_{_{L}}/(1+z)^2$ \cite{gfre}. The
luminosity distance can be determined by comparing the observed 
luminosity of an object to its known intrinsic luminosity: $4 \pi
d_{_{L}}^2=L/F$ where $F$ is the observed (measured) flux of light
received and $L$ is the object's intrinsic luminosity.
\subsubsection{Galaxy number counts}
Another observable of interest is the source number counts as a
function of the redshift. An observer at the vertex ``here and now''
will count on the light cone a number $dN$ of sources between
redshifts $z$ and $z+dz$ in a solid angle $d\Omega$. It follows that
\begin{equation}
\frac{dN}{dz}=f_c n(v_o,r)r^2 d\Omega\frac{dr}{dz}
\end{equation}
where $n(v_o,r)$ is the number density of sources and $f_c$
is a fractional number indicating the efficiency of the counts
(completeness) \cite{stoegerellis}. This number corrects for errors in
source selection and detection, see
e.g. \cite{stoegerellis,ellisperry}. For simplicity, we can assume
that the necessary adjustments for the dark matter can be incorporated
via $f_c$. The energy-density follows  
\begin{equation}
\rho(v_o,r)=n(v_o,r)\,M \label{counts}
\end{equation} 
where $M$ is the average rest mass for the counted sources.
\subsection{A fitting procedure algorithm}
As discussed earlier, the approach used allows us to integrate explicitly 
the model given observational data. As a first step, we rearrange the
model equations. We combine equations (\ref{dustrhomass}) evaluated at
$v=v_o$ with equation (\ref{dustredshift1}) and use $c(0,v_o)=1$ to
obtain    
\begin{equation}
\frac{c'(r,v_o)}{c^3(r,v_o)}=4 \pi (1+z)^2(r,v_o) \rho(r,v_o))
\end{equation}
which integrates to 
\begin{equation}
c(r,v_o)={\frac {{ 1}}{\sqrt {1 -8\,\pi \,\int \!\left ({1+z}
\right )^{2}(r,v_o)r\rho(r,v_o){dr}}}}.\label{integratedc1}
\end{equation}
Integrating (\ref{dustrhomass}) for $m(r,v_o)$ gives
\begin{equation}
m(r,v_o)={\frac{1}{2c(r,v_o)}}\Big{[}{\int{\Big{(}c'(r,v_o)}+4 \pi r
\rho(r,v_o) c(r,v_o)\Big{)}rdr}\Big{]}\label{integratedm1}
\end{equation}
where we also used $m(0,v_o)=0$.
Now, the observations provided as polynomial functions $\rho(z)$ and
$r(z)$ fitted to the data can be used to integrate explicitly for
$m(r,v)$ and $c(r,v)$. The steps for the fitting algorithm are as
follows:  
\begin{itemize}
\item{Express cosmological data as polynomial functions for:\\
i) The energy-density $\rho(r,v_o)$ from galaxy number counts. Many
projects are accumulating very large amounts of data. See, for example,
\cite{counts1} for the ``Sloan Digital Sky Survey''.\\ 
ii) The observer area distance $r(z,v_o)$ from ``standard candles''
projects in which it is possible to measure the redshift and the
distance independently. The accumulating data from the 
supernovae cosmology projects are very promising. See, for example,
\cite{supernovae1} for the ``High-Z SN Search'' project,
\cite{supernovae2} for the
``Supernova Cosmology Project'' and \cite{supernovae3} for ``The
Supernova Acceleration Probe (SNAP)'' project.}
\item{Invert the function $r(z,v_o)$ to obtain $z(r,v_o)$.}
\item{This can in turn be used to write the energy-density polynomial
  function as $\rho(z(r,v_o),v_o)$.} 
\item{Now, with $z(r,v_o)$ and $\rho(r,v_o)$ expressed as functions of
  $r$ (and not $z$), integrate equation (\ref{integratedc1})
  over $r$ to obtain $c(r,v_o)$ on the light cone.}
\item{With $c(r,v_o)$ determined, integrate equation
 (\ref{integratedm1}) over $r$ to obtain  $m(r,v_o)$ on the light cone.} 
\item{Finally, with $c(r,v_o)$ and $m(r,v_o)$ determined, use
  equations (\ref{dustconst1}) and (\ref{dustconst2}) } to integrate
  over v. 
\end{itemize}
  The level of difficulty of this last step can be monitored using the
  analytical forms used for $\rho(z)$ and $r(z)$ and remains
  a tractable problem while integrating the null geodesic equation
  in the standard 1+3 form of the LTB models is not tractable
  (see e.g. \cite{maartens}), and one has to recourse to numerical 
  integrations \cite{ribeiro}. 
  Moreover, the fitting procedure 
  has the interesting feature of incorporating the observations 
  in the process of integrating explicitly for the metric functions.

  It is worth mentioning that in principle
  the information on our light cone can not determine its  future
  evolution uniquely. We need to make the reasonable assumption that
  there will be no future events in the cosmic evolution
  that will  invalidate the entire data obtained from our light cone,
  see e.g. \cite{ellisstoeger}. 
Furthermore, one must keep in mind the usual limitation of the 
underlying models used here as they are spherically symmetric around 
our worldline and more general inhomogeneous models should be 
considered in future studies of fitting procedures.
\section{Summary}

We expressed inhomogeneous cosmological models in null spherical
non-comoving coordinates using the Bondi spherical metric. A known
difficulty in using inhomogeneous models is that the null geodesic
equation is not integrable in general. Our  choice of null coordinates
solves the radial null geodesic by construction. We identified the
general meaning of the metric function $c(r,v)$ to be the reciprocal
of the optical expansion. We used an approach where the velocity field
is uniquely calculated from the metric rather than put in by
hand. Conveniently, this allowed us to explore models in a
non-comoving frame of reference. In this frame, we find that the velocity
field has shear, acceleration and expansion rate in general. 
In this set of coordinates, we showed that a comoving frame 
is not compatible with expanding perfect fluid models
and dust models are simply not possible in this frame. 
We described then perfect fluid and dust models in a 
non-comoving frame. The framework developed allows one to
outline a fitting procedure where observational data
can be used directly to integrate explicitly for the models. 

\begin{acknowledgments}
The author thanks Kayll Lake and Roberto Sussman for useful discussions.
This work was supported by the Natural Sciences and Engineering Research 
Council of Canada (NSERC). Portions of this work were made possible by use 
of \textit{GRTensorII} \cite{grt}. 

\end{acknowledgments}

\begin{appendix}
\section{Null geodesic equation}\label{appNull}
The paths of light rays are described by null geodesic trajectories
under the eikonal assumption \cite{mtw}. The geodesic 
trajectories are determined by solving the null geodesic equation 
\begin{equation}
k_{\alpha;\beta}k^{\beta}=0\label{nullgeo}
\end{equation}
where $k^{\alpha}$ is a null vector ($k^{\alpha}k_{\alpha}=0$)
tangent to the null
geodesics, $k^{\alpha}=\frac{dx^{\alpha}}{d\lambda}$ where $\lambda$
is an affine parameter. For the Bondi metric (\ref{metric}), the 4
equations (\ref{nullgeo}) are all satisfied for $v=constant$. 
\section{Expressions for $G^{\alpha}_{\beta}$ components} \label{appEinstein}
The expressions for the components of the mixed Einstein tensor are as follows
\begin{equation}
{G^{r}_{r}=\frac{2c'}{cr}(1-\frac{2m}{r})-\frac{2m'}{r^2}}
\end{equation}
\begin{equation}
{G^{v}_{r}=\frac{2c'(r,v)}{c^2 r}}
\end{equation}
\begin{equation}
{G^{r}_{v}=\frac{2m^{\bullet}}{r^2}}
\end{equation}
\begin{equation}
{G^{v}_{v}=-\frac{2m'}{r^2}}
\end{equation}
\begin{eqnarray}
{G^{\theta}_{\theta}=G^{\phi}_{\phi}= \frac{c'}{cr}\Big{(}1-3m'+\frac{m}{r}-
\frac{c^{\bullet}r}{c^2} \Big{)}}+  &  \\
{\frac{c''}{c}\Big{(}1-\frac{2m}{r}\Big{)}-
\frac{m''}{r}+\frac{c^{\bullet}{'}}{c^2}} &
\end{eqnarray}
where $ ^{\bullet} \equiv \frac{\partial}{\partial v}$, $ ' \equiv
\frac{\partial}{\partial r}$, $c \equiv c(r,v)$ and $m \equiv
m(r,v)$. We note that these components are related by  
\begin{equation}
{G^{r}_{r}-G^{v}_{v}=G^{v}_{r}c(r,v)\Big{(}1-\frac{2m(r,v)}{r}\Big{)}.}
\label{simplify}
\end{equation}
\section{The Weyl tensor  and the
  condition for conformal flatness} \label{appWeyl}
The structure of the Weyl tensor $C_{\alpha\beta\gamma\delta}=0$ is
usually explored to derive the conformally flat case of a cosmological
solution (i.e. $C_{\alpha\beta\gamma\delta}=0$). This can reveal
the limits of the model's parameters for which it reduces to a
Lemaitre-Friedmann-Robertson-Walker (LFRW) model. The non-vanishing
components of the mixed Weyl tensor for the metric (\ref{metric}) are
related and given by 
\begin{eqnarray}
{C_{rv}{}^{rv}=C_{\theta\phi}{}^{\theta\phi}=2C_{r\theta}{}^{r\theta}=
2C_{r\phi}{}^{r\phi}= }& \nonumber \\ {2C_{v\theta}{}^{v\theta}=2C_{v\phi}{}^{v\phi}=
{{\cal{W}}{(r,v)}}} & \label{WeylComponents} 
\end{eqnarray}
where 
\begin{eqnarray}
\cal{W}}{(r,v)}  \equiv  
{\frac{1}{c^3 r^3}\Big{[}c^3 r(m''r-4m'+\frac{6m}{r})-c^2 c''r^3\;\;\;\;\; &\label{W}\\
{\Big{(}1-\frac{2m}{r}\Big{)} + c^2c'
r^2(1+3m'-\frac{5m}{r})+c^{\bullet}c'r^3-c^{\bullet}{'}cr^3\Big{]}}&
\nonumber 
\end{eqnarray}
The condition for conformal flatness of the models is therefore ${{\cal{W}}{(r,v)}=0}$.
\end{appendix}

\end{document}